\documentclass{elsart1p}
\usepackage{epsfig}
\usepackage{amssymb}
\usepackage{amsmath}
\usepackage{amsfonts} 
\usepackage[usenames]{color}

\journal{Physics Letters B}

\begin{document}

\begin{frontmatter}

\title{An improved determination of the two--nucleon induced non mesonic weak decay of $\Lambda$ hypernuclei}
\author[a,b]{M.~Agnello},
\author[c]{L.~Benussi}, \author[c]{M.~Bertani}, \author[d]{H.C.~Bhang},
\author[e,f]{G.~Bonomi}, \author[g,b]{E.~Botta},
\author[h]{M.~Bregant}, \author[g,b]{T.~Bressani}, \author[b]{S.~Bufalino\thanksref{cor1}},
\author[i,b]{L.~Busso}, \author[b]{D.~Calvo}, \author[j,k]{P.~Camerini},
\author[l]{B.~Dalena}, \author[g,b]{F.~De Mori}, \author[m,n]{G.~D'Erasmo},
\author[c]{F.L.~Fabbri}, \author[b]{A.~Feliciello}, \author[b]{A.~Filippi},
\author[m,n]{E.M.~Fiore}, \author[o]{H.~Fujioka},
\author[f]{P.~Genova}, \author[c]{P.~Gianotti}, \author[i]{N.~Grion},
\author[c]{V.~Lucherini}, \author[g,b]{S.~Marcello},
\author[p]{N.~Mirfakhrai}, \author[e,f]{F.~Moia}, \author[f,q]{P.~Montagna}, \author[r,b]{O.~Morra},
\author[o]{T.~Nagae},  \author[s]{H.~Outa}, \author[n]{A.~Pantaleo\thanksref{cor2}},
 \author[n]{V.~Paticchio}, \author[j]{S.~Piano},
\author[j,k]{R.~Rui}, \author[m,n]{G.~Simonetti},
\author[b]{R.~Wheadon}, \author[e,f]{A.~Zenoni}
\linebreak
\author{(The FINUDA Collaboration)}
\linebreak
and                                                                                                                   
\author[t,b]{G.~Garbarino}   
\address[a]{Dipartimento di Fisica, Politecnico di Torino, Corso Duca degli
Abruzzi 24, Torino, Italy}
\address[b]{INFN Sezione di Torino, via P. Giuria 1, Torino, Italy}
\address[c]{Laboratori Nazionali di Frascati dell'INFN, via E. Fermi 40,
Frascati, Italy}
\address[d]{Department of Physics, Seoul National University, 151-742 Seoul,
South Korea}
\address[e]{Dipartimento di Meccanica, Universit\`a di Brescia, via Valotti 9,
Brescia, Italy}
\address[f]{INFN Sezione di Pavia, via Bassi 6, Pavia, Italy}
\address[g]{Dipartimento di Fisica Sperimentale, Universit\`a di Torino,
Via P. Giuria 1, Torino, Italy}
\address[h] {SUBATECH, Ecole des Mines de Nantes, Universit\'e de Nantes, CNRS-IN2P3, Nantes, France}
\address[i]{Dipartimento di Fisica Generale, Universit\`a di Torino,
Via P. Giuria 1, Torino, Italy}
\address[j]{INFN Sezione di Trieste, via Valerio 2, Trieste, Italy}
\address[k]{Dipartimento di Fisica, Universit\`a di Trieste, via Valerio 2,
Trieste, Italy}
\address[l]{CERN, CH-1211 Geneva 23, Switzerland}
\address[m]{Dipartimento di Fisica Universit\`a di Bari, via Amendola 173,
Bari, Italy}
\address[n]{INFN Sezione di Bari, via Amendola 173, Bari, Italy}
\address[o]{Department of Physics, Kyoto University, Sakyo-ku, Kyoto Japan}
\address[p]{Department of Physics, Shahid Behesty University, 19834 Teheran, Iran}
\address[q]{Dipartimento di Fisica Nucleare e Teorica, Universit\`a di Pavia, via Bassi 6, Pavia, Italy}
\address[r]{INAF-IFSI, Sezione di Torino, Corso Fiume 4, Torino, Italy}
\address[s]{RIKEN, Wako, Saitama 351-0198, Japan}
\address[t]{Dipartimento di Fisica Teorica, Universit\`a di Torino, Via P. Giuria 1, Torino, Italy}                                                                                                    
\thanks[cor1]{Corresponding author: S. Bufalino, e-mail address: bufalino@to.infn.it}
\thanks[cor2]{deceased}

\begin{abstract}
The decay of $\Lambda$-hypernuclei without $\pi$ emission, known as Non Mesonic Weak Decay (NMWD), gives an effective tool to investigate $\Delta$S=1 four--baryon interactions. It was theoretically suggested that the two--nucleon induced mechanism could play a substantial role in reproducing the observed NMWD decay rates and nucleon spectra, but at present no direct evidence of such a mechanism has been obtained.
The FINUDA experiment, exploiting the possibility to detect both charged and neutral particles coming from the hypernucleus decay, has allowed us to deduce the relative weight of the two nucleon induced decay rate to the total NMWD rate. The value of $\Gamma_{2N}$/$\Gamma_{NMWD}$=0.24$\pm$${0.03_{stat}}^{+0.03_{sys}}_{{-{0.02_{sys}}}}$ has been deduced, with an error reduced by a factor more than two compared with the previous assessment. 
\end{abstract}
\begin{keyword}
$\Lambda$-hypernuclei \sep non mesonic weak decay \sep two nucleon induced decay
\PACS  21.80.+a \sep 25.80.Pw
\end{keyword}

\end{frontmatter}

\section{Introduction}
The Non Mesonic Weak Decay (NMWD) of Hypernuclei has stimulated a strong interest since the beginning of Hypernuclear Physics \cite{ches}. Indeed it has been realized that the two channels of NMWD,

\begin{equation}
^Z_{\Lambda}A \rightarrow ^{A-2}(Z-1)+n+p~
\label{eq1}
\end{equation}
and
 \begin{equation}
^Z_{\Lambda}A \rightarrow ^{A-2}Z+n+n~
\label{eq2}
\end{equation}
where A and Z indicate, respectively, the mass and atomic numbers of the decaying systems, are due to the occurrence of the two weak reactions:

\begin{equation}
\Lambda p \rightarrow np \quad(\Gamma_{p}) ~
\label{gammap}
\end{equation}
and
\begin{equation}
\Lambda n \rightarrow nn \quad(\Gamma_{n})~
\label{gamman}
\end{equation}
inside nuclei. These two decay processes are usually referred to as ``one--proton induced NMWD'' and, respectively, ``one--neutron induced NMWD'' of  $\Lambda$-Hypernuclei.
Reactions (\ref{gammap}) and (\ref{gamman}) constitute a unique class of four--baryon, strangeness non-conserving weak interactions and the determination of  their rates is of considerable interest. These rates cannot be determined experimentally by the direct reactions (\ref{gammap}) and (\ref{gamman}), due to the lack of suitable beams of Hyperons. The only way is to study them through their occurrence in Hypernuclei (reactions (\ref{eq1}) and (\ref{eq2})).\\
Unfortunately, also this last approach is not experimentally easy, due to the low rate of production of $\Lambda$--Hypernuclei in ground or excited states and to the detection efficiency of neutrons, which is small despite the large acceptance and good energy resolution apparatuses. For these reasons the experimental progress in this field has been limited for many years, in contrast to the vivid development of many theoretical approaches. Recent reviews can be found in \cite{alberico} and \cite{outa}. Ref.\cite{alberico2} first pointed out the possibility that a considerable amount of the strength for NMWD, up to about 20$\%$ of the total decay width, could be accounted for by the interaction of a $\Lambda$ with a pair of correlated nucleons in a nucleus, such as:

\begin{equation}
\Lambda np \rightarrow nnp \quad(\Gamma_{np})~,
\label{gammanp}
\end{equation}
\begin{equation}
\Lambda pp \rightarrow npp \quad(\Gamma_{pp})~,
\label{gammapp}
\end{equation}
\begin{equation}
\Lambda nn \rightarrow nnn \quad(\Gamma_{nn})~,
\label{gammann}
\end{equation}
referred to in the following as 2N-induced NMWD. The total non mesonic rate is given by $\Gamma_{\rm NMWD}$=$\Gamma_{1N} + \Gamma_{2N}$, with $\Gamma_{1N}$= $\Gamma_{n}+\Gamma_{p}$ and $\Gamma_{2N}$= $\Gamma_{np}+\Gamma_{pp}+\Gamma_{nn}$.\\ 
In addition to reaction (\ref{gammap}), (\ref{gamman}), (\ref{gammanp}), (\ref{gammapp}) and (\ref{gammann}), for light hypernuclei one can also study the so--called rare two--body decay. Recently FINUDA has been able to measure the decay yields and the branching ratios of the two-body decay channels ${\mathrm{^{4}_{\Lambda}He}}\rightarrow d+d$, ${\mathrm{^{4}_{\Lambda}He}} \rightarrow p+t$  and ${\mathrm{^{5}_{\Lambda}He}}\rightarrow d+t$ \cite{filippi}.
\\The suggestion of the existence of the 2N-induced NMWD was followed by detailed calculations thus stimulating a big experimental effort on the subject. A first summary of the theoretical issues can be found  in \cite{alberico}, whereas \cite{bauer2} (\cite{bauer}) reports the latest developments in the calculation of the NMWD widths (nucleon spectra).
First experimental determinations of $\Gamma_{2N}$ were done by indirect methods based on the fit of the ${\mathrm{^{12}_{\ \Lambda}C}}$ experimental inclusive proton spectra using IntraNuclear Cascade (INC) calculations including 2N-induced NMWD \cite{bhang} and reported for the $\Gamma_{2N}$/$\Gamma_{NMWD}$ ratio a value as large as 40$\%$.\\
Coincidence measurements of neutrons and protons, with an energy threshold of 30 MeV, following ${\mathrm{^{12}_{\ \Lambda}C}}$ NMWD were recently analyzed taking into account the angular correlations between the detected nucleons \cite{kim}. The coincidence spectra were analized using a new version of the INC code, with a strength varied to fit the $^{12}$C(p,p') total inelastic cross section, in order to account for the effect of the Final State Interaction (FSI) on the experimental spectra. From this analysis the experimental value $\Gamma_{2N}$/$\Gamma_{NMWD}$=0.29$\pm$0.13 was reported for ${\mathrm{^{12}_{\ \Lambda}C}}$ \cite{kim}.\\
A different approach to extract the strength for the 2N-induced NMWD was followed by the FINUDA Collaboration \cite{plb685}. Proton energy spectra of ${\mathrm{^{5}_{\Lambda}He}}$, ${\mathrm{^{7}_{\Lambda}Li}}$, ${\mathrm{^{9}_{\Lambda}Be}}$, ${\mathrm{^{11}_{\ \Lambda}B}}$,
 ${\mathrm{^{12}_{\ \Lambda}C}}$, ${\mathrm{^{13}_{\ \Lambda}C}}$, ${\mathrm{^{15}_{\ \Lambda}N}}$ and ${\mathrm{^{16}_{\ \Lambda}O}}$ were measured with good resolution ($\Delta$p/p=2\% FWHM for protons of 80 MeV) and with a kinetic energy detection threshold of 15 MeV. All the measured spectra showed a similar behaviour, i.e., a bump at about 80 MeV, roughly at the energy expected from reaction (\ref{eq1}). The bump is quite well defined in the high energy portion, whereas at low energies it is blurred in a continuum generated by FSI and superimposed to the 2N-induced NMWD contribution. Under very simple hypotheses, the contributions from FSI and 2N-induced NMWD were disentangled,  providing: $\Gamma_{2N}$/$\Gamma_{p}$=0.43$\pm$0.25 and $\Gamma_{2N}$/$\Gamma_{NMWD}$=0.24$\pm$0.10. \\
In this Letter we present an improvement of this method, made possible by the detection of both protons and neutrons, by which we determine the value of $\Gamma_{2N}$/$\Gamma_{NMWD}$ with an error reduced by a factor of more than two with respect to the previous FINUDA assessment \cite{plb685}.

\section{The experimental and analysis method}

The data were collected by the FINUDA experiment, installed at one of the two interaction regions of the DA$\Phi$NE ($e^{+},e^{-}$) $\phi$--factory  of Laboratori Nazionali di Frascati (INFN--Italy) and correspond to an integrated luminosity of about 1.2 $fb^{-1}$. A detailed description of the FINUDA experiment can be found
in ~\cite{fnd,fnd2}.

We do not report here experimental and analysis details already described in \cite{plb685} and references therein. Some information concerning the performances of the detectors which are relevant for the discussion of this new analysis are here recalled together with the features of the neutron detection.

FINUDA was a magnetic spectrometer, immersed in a uniform
solenoidal magnetic field of 1 T and optimized for the detection of charged particles, with an angular coverage of $\sim$ 2$\pi$ sr \cite{fnd}. The outer FINUDA detector, called TOFONE \cite{plb685,fnd} was a barrel of 72 plastic $(CH)_n$ scintillator slabs (255 cm long and 10 cm thick), used essentially for trigger and charged particle P.Id. (by Time Of Flight). It was also used to detect neutrons (and photons) with an efficiency of about 10$\%$ for neutrons in the kinetic energy range 15-150 MeV. \\The inner FINUDA detector was an hodoscope of 12 scintillator thin slabs (TOFINO) arranged around the beam pipe at the ($e^{+},e^{-}$) interaction point;
it was used for trigger purposes and to identify the charged kaons discriminating them from minimum ionizing (Bhabha or beam background) particles. The particle Time--Of-Flight was measured by the (TOFINO--TOFONE) system.\\
Neutrons and photons are identified looking for events in which TOFONE elements were not connected to the curved trajectories belonging to charged particles.

\begin{figure}[htbp]
\begin{center}
\includegraphics[width=90mm]{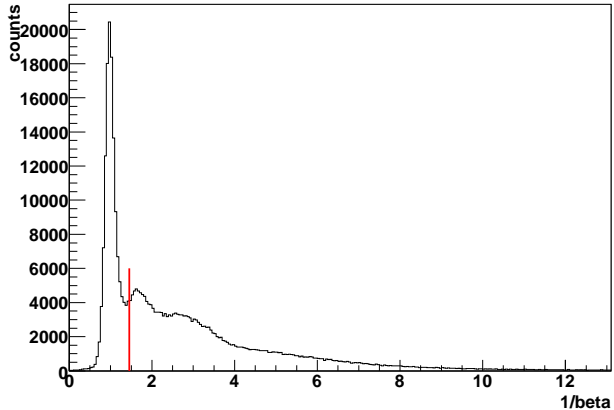}
\caption{Distribution of 1/$\beta$ for neutral candidates; the line indicates the cut applied in the data analysis.}
\label{figbeta}
\end{center}
\end{figure}

The $\beta$ of the neutral particle is evaluated by means of  $\beta$={\it (bof/tof)/c}, where {\it tof} is the measured time between the hit, on the TOFINO, of the K$^-$ from the $\phi$ decay just before stopping in the target (stopping time$\le$200 ps), and the hit on the TOFONE and {\it bof} is the corresponding  base of flight.\\
The analysis of the $\beta$ values of the neutral candidates allows us to discriminate neutrons from $\gamma$'s emitted, for istance, in the $\pi^{0}$ decay. Fig.\ref{figbeta} shows the distribution of 1/$\beta$ for selected neutral candidates. The peak centered at 1/$\beta$=1 is due to $\gamma$'s and it is followed by the contributions due to neutrons and to $\gamma$'s from  $\pi^{0}$ decay, which may be delayed if they are produced in the decay of kaons or hyperons. A contamination from charged particles back--emitted, following $\gamma$ interaction in the iron yoke of the FINUDA magnet and from neutron scattering in the FINUDA apparatus, is also present.\\
Applying a cut on the 1/$\beta$ value (1/$\beta$$\ge$1.47 as indicated by the line in Fig.\ref{figbeta}), it is possible to eliminate the $\gamma$ peak from the neutron candidates detection. As it will be discussed in the following the most effective way to identify neutrons from NMWD is to tag them when they are accompained by a coincidence proton. In the analysis we selected all the events in which a neutron and a proton are emitted in coincidence with a $\pi^{-}$ having  a momentum corresponding to the $\Lambda$--hypernucleus formation in its ground state or in a low lying excited one, decaying to the ground state by electromagnetic emission.

For the calibration of the neutron energy scale we used the monochromatic neutrons (185 MeV/c) produced in the decay at rest of the $\Sigma^+$ coming from the $K^-_{stop}+p \rightarrow\Sigma^++\pi^-$ reaction. The experimental peak is centered at 187.6 MeV/c with a $\sigma$=9.4 MeV/c.
The precision on the determination of the impact point of the neutrons on the scintillator slab is $\sigma$=6 cm; by combining this value with the timing resolution of  $\sigma$=780 ps we finally obtain for the overall energy resolution on the neutron $\sim$13$\%$ at 10 MeV and $\sim$20$\%$ at 100 MeV. A detailed description of the TOFONE performances can be found in \cite{tof}.

We analyzed neutron spectra coming from NMWD of Hypernuclei by using the same procedures adopted for the proton spectra described in \cite{bufalino}, i.e., by requiring the coincidence of a $\pi^{-}$ from the K$^{-}$ interaction vertex with a momentum compatible with the formation of the Hypernucleus with a bound $\Lambda$, but the result was unsuccessful. The neutron spectra were affected by a huge background, due to the contaminations described above, that we could not reduce neither by applying suitable cuts on the experimental spectra nor with the help of simulations.
We then considered the neutron spectra obtained by requiring not only the presence of a $\pi^{-}$, but also of a proton in quasi b.t.b. (back--to--back) correlation with the neutron (cos$\theta$(np)$\leq$-0.8). The number of such triple coincidence events was quite low, typically of the order of twenty for each nuclear target, and we could not infer from their distribution reliable conclusions. We added  the events from all hypernuclear species, and we compared neutron and proton spectra obtained after the acceptance correction and the subtraction of the background due to the $K^{−}(np) \rightarrow \Sigma^{-}p$ absorption, followed by the in–flight $\Sigma^{-}\rightarrow \pi^{-}n$ decay \cite{bufalino}. 

Fig.\ref{fig2} (first row) shows the result; the two spectra, 2a) for protons and 2b) for neutrons, are quite similar, as expected. 
The proton coincidence with a quasi b.t.b. neutron enhances the number of events due to the process (\ref{gammap}) with respect to the 2N-induced channels. On the other hand the quasi b.t.b requirement reduces the number of events in the low energy region, which are due to the channel (\ref{gammap}) and (\ref{gamman}) followed by a re--scattering inside the nucleus (FSI); in addition the b.t.b. correlation reduces the number of events from the 2N-induced decay (\ref{gammanp}), which are expected to exhibit a typical 3-body phase space angular and energy distribution. A confirmation of such expectation was found by requiring a tighter angular correlation  (cos$\theta$(np)$\leq$$-$0.9) and the result is reported in Fig.~\ref{fig2}c) for protons  and ~\ref{fig2}d) for neutrons:
\begin{figure}[htbp]
\begin{center}
\includegraphics[width=120mm]{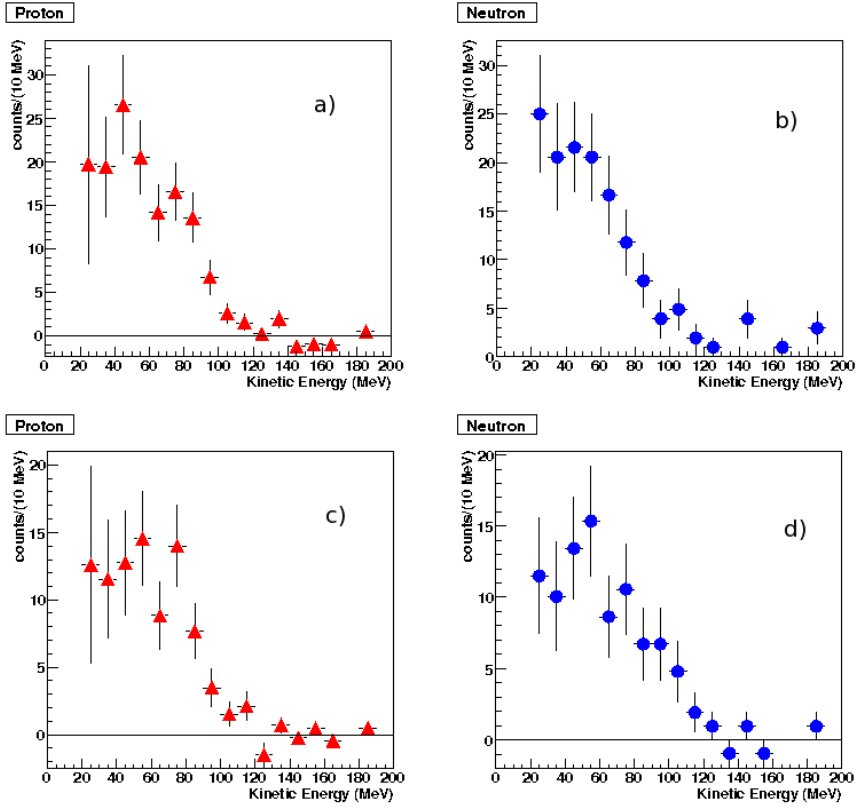}
\caption{Upper row: kinetic energy spectra of proton (a) and neutron (b) with a  quasi b.t.b. correlation (cos$\theta$(np)$\leq$-0.8) detected in coincidence with  a $\pi^{-}$ with a momentum compatible with the formation of a $\Lambda$ hypernucleus; lower row: kinetic energy spectra of proton (c) and neutron (d) with an angular correlation of (cos$\theta$(np)$\leq$$-$0.9) detected in coincidence with  a $\pi^{-}$ with a momentum compatible with the formation of a $\Lambda$ hypernucleus.}
\label{fig2}
\end{center}
\end{figure}
 the low energy region for both neutron and proton spectra is depleted and this confirms the validity of the analysis method.\\

Taking into account the efficacy of the method described above, we have analyzed the triple ($\pi^{-}$, $n$, $p$) coincidences considering the total number of events corresponding to each hypernucleus. 
We showed in \cite{plb685} that the single proton spectra from NMWD of ${\mathrm{^{5}_{\Lambda}He}}$ and the p-shell Hypernuclei ${\mathrm{^{7}_{\Lambda}Li}}$, ${\mathrm{^{9}_{\Lambda}Be}}$, ${\mathrm{^{11}_{\ \Lambda}B}}$, ${\mathrm{^{12}_{\ \Lambda}C}}$, ${\mathrm{^{13}_{\ \Lambda}C}}$, ${\mathrm{^{15}_{\ \Lambda}N}}$ and  ${\mathrm{^{16}_{\ \Lambda}O}}$ had a similar behaviour, i.e, a bump around 80 MeV, due to reaction (\ref{gammap}) without FSI, and  a rise in the low energy region, increasing with A, due to FSI and to channel (\ref{gammanp}). A fit to each spectrum beyond 80 MeV by using a Gaussian function with free mean values and sigma allowed us to disentangle the contributions due to process (\ref{gammap}), the effect of FSI and process (\ref{gammanp}) and to determine: $\Gamma_{2N}$/$\Gamma_{p}$=0.43$\pm$0.25 and $\Gamma_{2N}$/$\Gamma_{NMWD}$=0.24$\pm$0.10 \cite{plb685}.\\
In the analysis of the events due to triple ($\pi^{-}$, $n$, $p$) coincidences first of all we fixed for each hypernucleus a proton kinetic energy limit $E_{p}$ placed at 20 MeV below the mean value $\mu$ of the Gaussian found in \cite{plb685}. In order to enhance the contribution of the 2N--induced NMWD we also chose an upper limit for the proton--neutron angular correlation of cos$\theta$(np)$=$$-$0.8.\\
We classified then the triple coincidence ($\pi^{-}$, $n$, $p$) events into four groups, according to the following criteria:

\begin{enumerate}
\item
events with proton kinetic energy $E_{p}$ larger than the limit and cos$\theta$(np)$\leq$ $-$0.8. These events should stem  mainly from the process (\ref{gammap}) without FSI of the proton, even if a negligible contribution from the reaction (\ref{gammanp}) is expected. Note that we neglect the process (\ref{gammapp}), since $\Gamma_{pp}\simeq 0.15\Gamma_{np}$ \cite{nucl-th};
\item
events with $E_{p}$ larger than the limit and cos$\theta$(np)$\geq$ $-$0.8. These events should correspond mainly to the process (\ref{gammap}) or (\ref{gamman}) followed by a FSI in addition to a small contribution from (\ref{gammanp}). 
\item
events with $E_{p}$ lower than the limit and cos$\theta$(np)$\le$ $-$0.8. These events should correspond
mainly  to the process (\ref{gammap}) with FSI of the proton. 
\item
events with $E_{p}$  lower than the limit and cos$\theta$(np)$\geq$ $-$0.8. These events should correspond mainly to the process $\Lambda np\to nnp$ with a small contribution of FSI. 
\end{enumerate}
The cuts applied in the selection (iv) produce an underestimation of the population of this group which can be estimated to be $20\%$. This estimation is based on the assumption of a  $\geq$4--body phase space distribution of cos$\theta$(np); for the understimation due to the cut on $E_{p}$ the calculation of \cite{prc69} has been used.\\
The neutron acceptance for these events was evaluated by taking into account the apparatus geometry, the efficiency of the FINUDA pattern recognition algorithm and the quality cuts applied to the real data. The emission of a neutron from a $K^{-}$ stopped in the target was simulated, assigning to the neutron, emitted in coincidence with a proton and a negative pion,  a flat momentum distribution from 100 MeV/c to 700 MeV/c. The proton momentum was also simulated with a flat distribution in the same range of the neutron  and the $\pi^{-}$ momentum was simulated in the range (270--290) MeV/c corresponding to the formation of the $\Lambda$ hypernuclei. The acceptance function was evaluated target by target in different apparatus sectors and it was found to be flat within 20$\%$ in the energy range (10--100) MeV. 

\section{Experimental results and discussion}
\begin{figure}[htbp]
\begin{center}
\includegraphics[width=90mm]{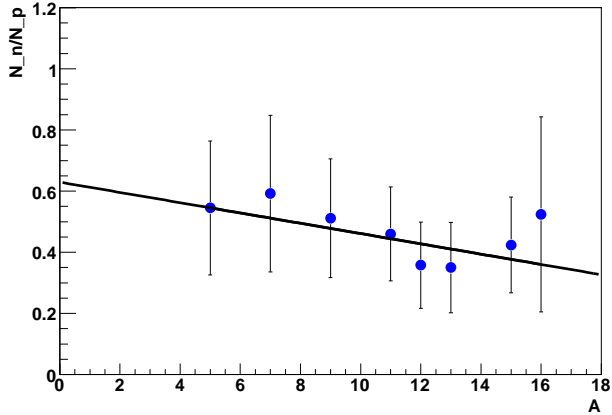}
\caption{The ratio $N_{n}(E_p\leq (\mu-20MeV), cos\theta(np)\ge -0.8)$/$N_{p}(E_p >\mu)$ as a function of the mass number A.}
\label{fig5}
\end{center}
\end{figure}

We considered the dependence on the mass number A of the ratio R between the events of the selection (iv) and the number of protons with $E_{p}$$\geq$$\mu$. This last number was evaluated by integrating the number of events of the proton spectra above the mean value $\mu$ of the Gaussian fit of \cite{plb685} for each hypernucleus.
We have:
\begin{eqnarray}
\label{vs-a}
R \equiv \frac{N_{n}(E_p\leq(\mu-20MeV), cos\theta(np)\ge-0.8)}{N_{p}(E_p > \mu)}=  \nonumber \\
  \frac{0.8 N(\Lambda np\to nnp)+ N_{n}^{FSI_{1N}}+N_{n}^{FSI_{2N}}}{0.5N(\Lambda p\to np)+ N_{p}^{FSI_{1N}}}~
\end{eqnarray}
where  $N_{n}(E_p\leq (\mu-20MeV), cos\theta(np)\ge$-$0.8)$ corresponds to the number of neutrons fulfilling the condition (iv) and that we estimate to be $80\%$ of the total number of neutron coming from the process (\ref{gammanp}), $N_{p}(E_p >\mu)$ is the number of protons with an energy higher than the mean value of the fit reported in \cite{plb685}; $N_{n}^{FSI_{1N}}$, $N_{p}^{FSI_{1N}}$, $N_{n}^{FSI_{2N}}$ are
 the number of protons or neutrons which suffer the final state interaction effect following the one--(1N) or two--nucleon (2N) induced NMWD.
In Fig.~\ref{fig5} the experimental values of this ratio for each hypernucleus are plotted as a function of $A$.\\
Due to FSI both $N_{n}(E_p\leq (\mu-20MeV), cos\theta(np)\ge -0.8)$ and $N_{p}(E_p > \mu)$ in (\ref{vs-a}) are expected to be proportional to A; we performed a fit with a function $R(A)=(a'+b'A)/(c+dA)$ and we found {\it d/c$\leq$10$^{-3}$}; so a simple linear fit $R(A)=(a+bA)$ can be applied. The result of such a fit ($a=0.63\pm 0.25$, $b=-0.017\pm 0.021$ and $\chi^{2}/ndf=0.298/6$) is reported in Fig.~\ref{fig5}.\\
We can make the assumption that $\Gamma_{np}$/$\Gamma_{p}$ is constant in the mass number range of the analyzed hypernuclei \cite{bauer2}. Moreover, one has:
\begin{equation}
\label{number-gamma}
\frac{N(\Lambda np\to nnp)}{N(\Lambda p\to np)}=\frac{\Gamma_{np}}{\Gamma_{p}},~
\end{equation}
thus eq.(\ref{vs-a}) provides:
\begin{equation}
\label{RA}
R(A)\equiv \frac{0.8\Gamma_{np}}{0.5\Gamma_{p}} + bA
\end{equation}
where {\it b} is taken from the fit and represents the small FSI contribution.
Eq.(\ref{RA}) can be solved for $\Gamma_{np}/\Gamma_p$ for each hypernucleus, giving:
\begin{equation}
 \frac{\Gamma_{np}}{\Gamma_{p}}=\frac{[R(A)-bA]}{1.6}~.
\label{a5}
\end{equation}
The values obtained for $\Gamma_{np}/\Gamma_{p}$ for the examined hypernuclei
are all compatible with each other within errors. The final result can thus be given by
the weighted average over all the available nuclear species:
\begin{equation}
 \frac{\Gamma_{np}}{\Gamma_{p}}=0.39\pm {0.07_{stat}}^{+0.04_{sys}}_{{-{0.03_{sys}}}}~.
\label{a6}
\end{equation}

The quoted statistical error is less than half the one obtained in the proton spectra analysis of \cite{plb685}. This is because, unlike \cite{plb685}, using the described selection criteria, the ratios R and $\Gamma_{np}/\Gamma_{p}$ are linearly related. The experimental inaccuracies due to the background subtraction and the efficiency correction are included in the statistical error. The systematic errors instead take into account the effect of slight variations in the selection criteria applied on cos$\theta$(np) and on the limit of the proton energy $E_{p}$; the variation due to the use of the weighted average instead of the normal one is also included. In addition, the possible uncertainty on the understimation of population of group (iv) has been included.\\ 
As pointed out in \cite{plb685}, the 2N--induced NMWD can be assumed to be dominated by the
$\Lambda np\to nnp$ channel and
the recent microscopical calculation of \cite{nucl-th} delivers
$\Gamma_{np}:\Gamma_{pp}:\Gamma_{nn}=0.83:0.12:0.04$.  Ref.\cite{nucl-th} presents an improvement of the first calculation of the 2N-induced decay rates  performed within a nuclear matter framework \cite{alberico2}, where a  phenomenological description of the two--particle two--hole polarization propagator was adopted; the effects of Pauli exchange terms in the two--nucleon stimulated NMWD are also taken into account in \cite{nucl-th}.
Considering that the total error on $\Gamma_{np}/\Gamma_{p}$ quoted in eq.(\ref{a6}) is comparable with the contribution of ($\Gamma_{pp}+\Gamma_{nn}$), in the present analysis we cannot assume  $\Gamma_{2N}\sim\Gamma_{np}$ as in \cite{plb685}. 
We can determine $\Gamma_{2N}$/$\Gamma_{\rm NMWD}$ by adopting the same method employed in \cite{plb685}; using the experimental value of $\Gamma_{n}/\Gamma_{p}$ reported in \cite{bhang} for $\mathrm{^5_{\Lambda}He}$ and  $\mathrm{^{12}_{\ \Lambda}C}$, our determination of $\Gamma_{np}/\Gamma_{p}$ and taking into account that  $\Gamma_{np}/\Gamma_{NMWD} \simeq 0.83(\Gamma_{2N}/\Gamma_{NMWD}$) \cite{nucl-th} we obtain:
\begin{equation}
 \frac{\Gamma_{2N}}{\Gamma_{\rm NMWD}}=\frac{\Gamma_{2N}/\Gamma_{p}}{(\Gamma_{n}/\Gamma_{p})+1 +(\Gamma_{2N}/\Gamma_{p})}=0.24\pm {0.03_{stat}}^{+0.03_{sys}}_{{-{0.02_{sys}}}}~.
\label{a8} 
\end{equation}
This value supports the latest theoretical predictions \cite{bauer2} ($\Gamma_{2}/\Gamma_{\rm NMWD}$=0.26), the recent experimental results of \cite{kim} ($0.29\pm0.13$) and the previous FINUDA result \cite{plb685}, but bears a smaller error.

\section {Conclusions}
We performed a model--independent analysis of the
2N--induced NMWD process $\Lambda np \rightarrow nnp$ by analyzing the ($\pi^-$, $p$, $n$) triple coincidence events from K$^{-}$ stopped in thin nuclear targets with the FINUDA spectrometer at DA$\Phi$NE. The measurement of the $\pi^{-}$ momentum allowed the selection of events  coming from the decay of p-shell $\Lambda$-hypernuclei.
By applying appropriate cuts on the energies and angles of the protons and the neutrons we could identify events due mainly to $\Lambda np \rightarrow nnp$ NMWD and determine $\Gamma_{np}$/$\Gamma_{p}$=0.39$\pm$${0.07_{stat}}^{+0.04_{sys}}_{{-{0.03_{sys}}}}$. We also extracted $\Gamma_{2N}$/$\Gamma_{NMWD}$=0.24$\pm$${0.03_{stat}}^{+0.03_{sys}}_{{-{0.02_{sys}}}}$ reducing the total error by a factor of more than two with respect to the value published in \cite{plb685}. The obtained results are in agreement, within the errors, with previous experimental determinations, model dependent or not, and with theoretical calculations. We hope that these limited statistics results  will be improved by the experiments planned at J-PARC.\\

We dedicate this paper to the memory of Dr. Ambrogio Pantaleo passed away prematurely on June 23rd, 2010. Dr. Pantaleo participated very actively since the beginning to the FINUDA experiment, taking particular care of the neutron detection, on which the present analysis is based.

\section {Acknowledgements}

We acknowledge the European Community--Research Infrastructure Integrating Activity Study of Strongly Interacting Matter
(HadronPhysics2, Grant Agreement n. 227431; SPHERE network) under the Seventh Framework Programme of EU for the partial support of this work.


\begin{thebibliography}{00}

\bibitem{ches}
W.~Cheston and H.~Primakoff, {\it Phys. Rev.}  {\bf 92} (1953) 1537.

\bibitem{alberico}
W.M.~Alberico and G.~Garbarino, {\it Phys. Rep.} {\bf 369} (2002) p. 1; \\ in {\it Hadron Physics} (IOS Press, Amsterdam, 2005). Proc. of the International School of Physics ''E. Fermi'' Course CLVIII in Varenna (2004), edited by T. Bressani, A. Filippi and  U. Wiedner, p. 125.

\bibitem{outa}
 H.~Outa, in {\it Hadron Physics} (IOS Press, Amsterdam, 2005) p. 219. Proc. of the International School of Physics ''E. Fermi'' Course CLVIII in Varenna (2004), edited by T. Bressani, A. Filippi and  U. Wiedner.

\bibitem{alberico2}
W.M.~ Alberico, A.~ De Pace, M.~ Ericson and A.~ Molinari, {\it Phys. Lett.} {\bf B 256} (1991) 134.

\bibitem{filippi}
M. Agnello {\it et. al.}, 'Study of some two-body non-mesonic decays of $^4_\Lambda$He and $^5_\Lambda$He'.
e-Print: arXiv:1010.5616 [nucl-ex] (Oct. 2010).

\bibitem{bauer2}
E.~Bauer and G.~Garbarino, {\it Phys. Rev} {\bf C 81} (2010) 064315.

\bibitem{bauer}
E.~Bauer, G.~Garbarino, A.~Parreno and A.~Ramos, {\it Nucl. Phys.} {\bf A 836} (2010) 199.

\bibitem{bhang}
H.~Bhang  {\it et al.}, {\it Eur. Phys. J.} {\bf A 33} (2007) 259.


\bibitem{kim}
M.~Kim {\it et al.}, {\it Phys. Rev. Lett.} {\bf 103} (2009) 182502

\bibitem{plb685}
M.~Agnello, {\it et al.,} {\it Phys. Lett.} {\bf B 685} (2010) 247.

\bibitem{fnd}
M.~Agnello  {\it et al.,} {\it Nucl. Instrum. Meth.} {\bf A 570} (2007) 205.

\bibitem{fnd2}
M.~Agnello  {\it et al.,} {\it Phys. Lett.} {\bf B 622} (2005) 35.

\bibitem{tof}
A.~Pantaleo, {\it et al.,} {\it Nucl. Instr. Meth.} {\bf A 545} (2005) 593.

\bibitem{bufalino}
M.~Agnello, {\it et al.,} {\it Nucl. Phys.} {\bf A 804} (2008) 151.

\bibitem{nucl-th}
E.~Bauer and G.~Garbarino, {\it Nucl. Phys.} {\bf A 828} (2009) 29.

\bibitem{prc69}
 G.~Garbarino, A.~Parreno and A.~Ramos, {\it Phys. Rev.} {\bf C 69} (2004) 054603.



\end{thebibliography}
\end{document}